\documentclass[twocolumn,amsmath,amssymb,prl]{revtex4}

\usepackage{graphicx} 
\usepackage{dcolumn}  
\usepackage{bm}       

\begin{document}
\draft

\title{Correlated electrons in Fe-As compounds: a quantum chemical perspective}

\author{L. Hozoi and P. Fulde}
\address{Max-Planck-Institut f\"{u}r Physik komplexer Systeme,
         N\"{o}thnitzer Str.~38, 01187 Dresden, Germany}

\date\today

\begin{abstract}

State-of-the-art quantum chemical methods are applied to the study of
the multiorbital correlated electronic structure of a Fe-As compound, the 
recently discovered LiFeAs.
Our calculations predict a high-spin, $S\!=\!2$, ground-state configuration for
the Fe ions, which shows that the on-site Coulomb interactions are substantial. 
Also, orbital degeneracy in the $(xz,yz)$ sector and a three-quarter filling of these
levels suggest the presence of strong fluctuations and are compatible with a
low metallic conductivity in the normal state.
The lowest electron-removal states have As $4p$ character, in analogy to the 
ligand hole states in $p$-type cuprate superconductors.

\end{abstract}

\maketitle

Electron correlations in transition-metal (TM) solid-state compounds give rise to
a variety of less conventional phenomena.
Superconductivity in layered copper oxides at temperatures as high
as 100 K, for example, and the pairing mechanism in these systems are believed
to involve strong correlation effects among the valence Cu $3d$ electrons.
The discovery of the Fe$^{2+}$--\,As$^{3-}$ superconducting compounds \cite{feas_kamihara_08}
is the latest surprise in the field of $d$-electron systems.
Also in this case, the experimental data indicate unconventional superconductivity.
``Bad-metal'' conductivity in the normal state, a small carrier density,
a relatively small in-plane coherence length, and Uemura scaling in the
muon spin relaxation spectra \cite{feas_luetkens_08} are all reminiscent of
cuprate superconductors.

How strong correlations are in iron pnictides is an important issue.
In a first approximation, the charge distribution within the Fe $3d$ 
levels and the spin state depend on the intra-orbital Coulomb interaction, the
so-called Hubbard $U$. 
A $U$ value much larger than the crystal-field splittings will favor a
high-spin arrangement of the six Fe $3d$ electrons, while a low-spin, closed-shell
configuration is expected for small values of $U$.
The picture complicates when inter-orbital Coulomb and exchange interactions
are considered.
Additionally, correlation effects related to ligand $p$ to TM $d$ charge
transfer excitations may be important too, as discussed for the case of
another $3d^{6}$ system, the cobalt oxide perovskite LaCoO$_3$ \cite{LaCoO_hozoi_08}.

We investigate here the electronic structure of a Fe-As compound, the recently
discovered LiFeAs \cite{lfa_tapp_08}.
{\it Ab initio}, wave-function-based methods from modern quantum chemistry
are used in our study. 
We characterize the ground-state electron configuration for the undoped case
and provide new insight into the nature of doped carriers.
Our results lend credence to the view \cite{feas_haule_08,feas_craco_08} that
correlations are moderate to strong in Fe pnictides.

LiFeAs has a tetragonal crystal sructure, with the P4/$nmm$ space group
\cite{lfa_tapp_08}.
Different from other Fe-As compounds, it exhibits superconductivity at 
ambient pressure without chemical doping, with $T_c\!\approx\!18$ K. 
The common feature of the Fe pnictide superconductors is the Fe$_2$As$_2$
network of FeAs$_4$ tetrahedra.
Nearest neighbor (NN) FeAs$_4$ units share edges, while next-nearest-neighbor 
(NNN) tetrahedra share their corners.
In LiFeAs, the Fe$_2$As$_2$ layers are separated from each other by double
layers of Li ions.

The first step in our study is a ground-state restricted-Hartree-Fock (RHF)
calculation for the periodic crystal.
This calculation was carried out with the {\sc crystal} package \cite{crystal}.
We employed the lattice parameters reported in Ref.~\cite{lfa_tapp_08} and 
Gaussian-type, all-electron basis sets.
Basis sets of double-zeta quality from Towler's {\sc crystal} data basis
were applied for the As and Li ions \cite{BS_Towler,BS_As}.
For Fe, we used a basis set of triple-zeta quality, with $s$ and $p$ functions 
from Towler's data basis \cite{BS_Towler} and the $d$ functions developed by Seijo
{\it  et al.} \cite{BS_Fed_Seijo}.

The periodic RHF calculation yields a finite gap at the Fermi level.
For the RHF wave function, the $xz$ and $yz$ components are the highest among
the Fe $3d$ levels and unoccupied; the other Fe $3d$ orbitals are doubly
occupied.
We employ a reference system having the $x$ and $y$ axes rotated by $45^{\circ}$ 
with respect to the $a$ and $b$ coordinates of the P4/$nmm$ space group,
such that the As NN's of a given Fe site are situated either in the
$xz$ or $yz$ plane.

On-site and inter-site correlation effects are investigated in direct space
by means of multiconfiguration complete-active-space self-consistent-field
(CASSCF) and multireference configuration-interaction (MRCI) calculations.
The CASSCF wave function is constructed as a full configuration-interaction (CI)
expansion within a limited set of ``active'' orbitals \cite{QC_book_00}, i.e.,
all possible occupations are allowed for these active orbitals.
In the present study, the active orbital set contains all $3d$ orbitals at a 
given number of Fe sites.
Not only the CI coefficients but also the orbitals are variationally optimized
in CASSCF, which makes this method quite flexible.
MRCI wave functions are further constructed by adding single and double
excitations from the Fe $3s$, $3p$, $3d$ and As $4s$, $4p$ orbitals on top of the
reference CASSCF wave function, which is referred to as SD-MRCI \cite{QC_book_00}.

The quantum chemical computations are performed on a finite cluster
$\mathcal{C}$ including nine FeAs$_4$ tetrahedra -- a ``central'' FeAs$_4$ unit
plus four NN and four NNN tetrahedra -- and 16 Li neighbors of the As ions of
the ``central'' unit.
The orbital basis entering the correlation treatment is a set of {\it projected}
RHF Wannier functions:
localized Wannier orbitals (WO's) are first obtained with the Wannier-Boys localization
module \cite{zicovich_01} of the {\sc crystal} package and subsequently projected 
onto the set of Gaussian basis functions associated with the atomic sites
of $\mathcal{C}$ \cite{QPbands_0607}. 
Projected As $4p_x$ and Fe $3d_{xz}$ WO's, for example, are plotted in Fig.~1.
Moreover, the RHF data is used to generate an embedding potential for the
nine-tetrahedra fragment $\mathcal{C}$.
This potential is obtained from the Fock operator in the RHF calculation
\cite{QPbands_0607} and models the surroundings of the finite cluster, i.e., the
remaining of the crystalline lattice.
It is added to the one-electron Hamiltonian in the subsequent CASSCF/MRCI calculations
via an interface program developed in our laboratory \cite{crystal_molpro_int}. 
The CASSCF and MRCI investigations are carried out with the {\sc molpro}
program \cite{molpro_2006}.

\begin{figure}[b]
\includegraphics*[angle=0,width=0.80\columnwidth]{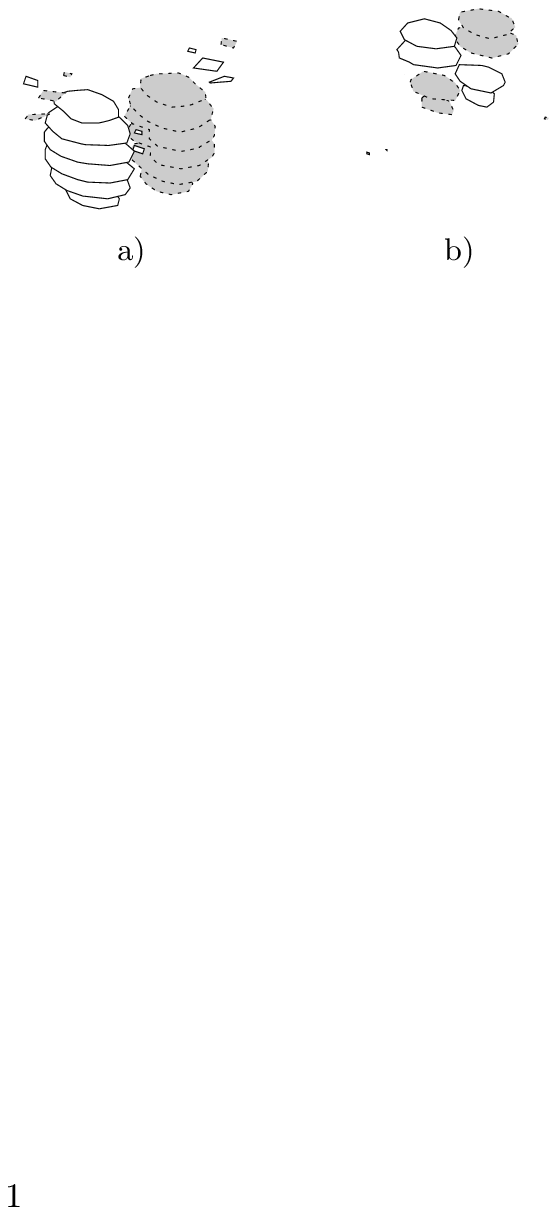}
\caption{
a) As $4p_x$ and b) Fe $3d_{xz}$ RHF WO's after projection onto the finite cluster,
see text.
The tails at nearby sites are very small.
For the Fe $3d_{xz}$ and $3d_{yz}$ WO's, the lobes pointing towards the
NN As sites are less extended.
}
\end{figure}

In a first set of CASSCF calculations, a number of nine sites are explicitly
correlated, those of the reference FeAs$_4$ tetrahedron and the four NN Fe ions.
This group of nine sites form the active region of the cluster, which we denote
as $\mathcal{C}_A$.
The other ions in $\mathcal{C}$, i.e., four Fe NNN's, 12 As, and 16 Li sites, form
a buffer region $\mathcal{C}_B$, whose role is to ensure an accurate representation
of the tails of the WO's centered in the active part $\mathcal{C}_A$.
For our choice of $\mathcal{C}_B$, the norms of the projected WO's at the central 
tetrahedron and NN plus NNN Fe sites are not lower than $99.5\%$ of the original
crystal WO's. 
While the occupied orbitals in the buffer zone are frozen, orbitals centered at sites
in the $\mathcal{C}_A$ region (and their tails in $\mathcal{C}_B$) are allowed to 
relax and polarize in the CASSCF study.  
For this first set of CASSCF calculations, the active orbital space consists of 25
Fe $3d$ orbitals, five at each active Fe site. 
Due to the large size of our CAS, i.e., 30 electrons in 25 orbitals, we restrict
our calculations to the case of high-spin (i.e., ferromagnetic) couplings
among neighboring Fe ions, although the experiments indicate antiferromagnetic 
inter-site interactions.
It is known, however, that the local charge distribution at a given TM
site does not depend on the nature of the inter-site $d$-$d$ magnetic couplings.
We observed this in the case of the Cu oxide superconducting compounds, for
example.
Also, a closed-shell representation of the Fe ions in the rest of the crystal
is acceptable, given the fact that the Fe NN's are treated by CASSCF,
on the same footing with the reference Fe site.

The CASSCF calculations show that at each Fe site the $3d$ electrons are
coupled into quintet states.
In contrast to the RHF results \cite{footnote_RHFgs}, the $xz$ and $yz$ components 
are the lowest in energy, such that the sixth electron is accomodated into these
levels.
The ground-state wave function is thus doubly degenerate,
$d_{xz,yz}^3d_{xy}^1d_{3z^2-r^2}^1d_{x^2-y^2}^1$.
The first excited state is also a quintet, at 0.25 eV higher energy, and
corresponds to a $(xz,yz)\!\rightarrow\!xy$ transition \cite{footnote_ddHS}.

The fact that the relative energies of the Fe $3d$ levels are not consistent with
simple considerations based on ligand-field theory for tetrahedral coordination,
which predicts that the $e_g$ levels are lower than the $t_{2g}$ states,
was pointed out before, see, e.g., Refs.~\cite{feas_haule_08,feas_boeri_08}.
This seems to be related to the distortion of the As tetrahedra, which are squeezed
in the $z$ direction, and the presence of direct, nearest-neigbor Fe $d$-$d$
orbital overlap \cite{feas_haule_08,feas_boeri_08}.

Our finding of a high-spin (HS), $S\!=\!2$ ground-state configuration agrees with 
the results of dynamical mean-field theory (DMFT \cite{dmft_rev_GK06}) investigations 
in the moderate to strong coupling regime by Haule {\it et al.} \cite{feas_haule_08}
and Craco {\it et al.} \cite{feas_craco_08}.
Also, a three-quarter filling of the degenerate $d_{xz}$ and $d_{yz}$ bands as
found in our calculations is compatible with the low metallic conductivity in the
normal state of these systems.
Systems where a pair of degenerate orbitals accomodates one electron or one hole
often display very rich physics, involving couplings among the charge, lattice,
and spin degrees of freedom. 
The structural transition at about 150 K in some of the Fe-As compounds
\cite{feas_delaCruz_08,feas_rotter_08} might occur
such that the degeneracy of the $xz$ and $yz$ orbitals is lifted.
This issue remains to be investigated in future work. 

It would be instructive to determine the relative energies of states involving
low-spin couplings at a given Fe site. 
However, for a CAS with five Fe ions and 25 orbitals, such investigations
are quite difficult.
Since the on-site interactions are much larger than the inter-site $d$-$d$
spin couplings, the states related to low-spin configurations at a given site
are among the highest in a multitude of low-spin excited states.
Identifying and optimizing those states is a very tedious task.
In order to access those states, we reduce then the orbital space in the CASSCF
calculations to the set of five $3d$ functions at the central Fe site.
For each of the Fe NN's, the $3d$ electrons are ``forced'' into a $t_{2g}^6$
closed-shell configuration. 

With this choice of the CAS, the lowest intermediate-spin (IS), $S\!=\!1$
and low-spin (LS), $S\!=\!0$ states require excitation energies of 1.91
and 2.34 eV, respectively, with respect to the HS ground-state (see
Table I).
The lowest $S\!=\!0$ state corresponds to a $t_{2g}^6$ orbital occupation
and the lowest $S\!=\!1$ state corresponds to a $t_{2g}^5e_{g}^1$ configuration,
more precisely $d_{xy}^2d_{xz,yz}^3d_{3z^2-r^2}^1$. 
 
On top of the CASSCF wave functions, we further performed MRCI
calculations with single and double excitations from
the Fe $3s$, $3p$, $3d$ and As $4s$, $4p$ orbitals at the central tetrahedron.
The SD-MRCI treatment decreases the HS-IS and HS-LS energy splittings to 1.30
and 1.74 eV, respectively (lowest line in Table I).
Now, the IS-LS energy difference, for example, can be used to extract
information on the magnitude of parameters such as the intra-orbital and
inter-orbital Coulomb interactions $U$ and $U'$ and the inter-orbital
exchange coupling $J_H$.
The $t_{2g}^5e_{g}^1\!-\!t_{2g}^6$ energy difference can be expressed as
$\Delta E = U + J_H - U' - \Delta_{\mathrm{CF}}$.
For the sake of simplicity, we assume that the Coulomb repulsion terms between
different $t_{2g}$ and $e_g$ orbitals have all the same value.
A unique value is also assumed for the $t_{2g}\!-\!e_g$ inter-orbital exchange
couplings \cite{footnote_UJterms}.
To determine the crystal-field splitting $\Delta_{\mathrm{CF}}$ between the ($xz$,$yz$)
and $3z^2\!-\!r^2$ components, we perform SD-MRCI calculations for the 
($xz$,$yz$)\,$\rightarrow$\,$3z^2\!-\!r^2$ excitation energy in the HS, $S\!=\!2$ 
configuration.
By SD-MRCI, this splitting is 0.71 eV.
Since $\Delta E\!=\!0.44$ eV, see the lowest line in Table I, it follows that
$U+J_H-U'\!=\!1.15$.
Using the relation $U'\!=\!U\!-\!2J_H$, we find $U\!-\!U'\!\approx\!0.8$ eV and
$J_H\!\approx\!0.4$ eV.

\begin{table}[t]
\caption{
Relative energies for the HS, IS, and LS configurations
by CASSCF and SD-MRCI calculations, see text.
The HS state is always the lowest.
The energy of the HS state in the SD-MRCI calculation was taken as
reference.
}
\begin{ruledtabular}
\begin{tabular}{lrrr}
Relative energy (eV)       &HS      &IS      &LS     \\

\colrule
\\
CASSCF                     &13.16   &15.07   &15.50  \\
SD-MRCI; Fe $3s$,$3p$,$3d$,
As $4s$,$4p$               &0       &1.30    &1.74   \\
\end{tabular}
\end{ruledtabular}
\end{table}

As expected, $U\!-\!U'$ does not depend on the electron configuration at the NN
Fe sites: if in the CASSCF and SD-MRCI calculations for the central
tetrahedron we adopt a $e_g^4d_{xy}^2$ configuration at the NN Fe sites, as
found in the periodic RHF calculation \cite{footnote_RHFgs}, the difference
between $U$ and $U'$ remains 0.8 eV.
Our results provide a lower limit for the value of the Hubbard $U$.
We note that estimates based on density-functional (DF) calculations
strongly depend on the type and size of the Wannier-like orbital basis.
Constrained DF computations by Anisimov {\it et al.} \cite{feas_anisimov_U}
using a WO basis restricted to the Fe $3d$ orbitals yield $U\!=\!0.55$
and $J_H\!=\!0.5$ eV.
With an extended orbital basis including As $4p$ functions, it was found that
$U\!=\!3\!\div\!4$ and $J_H\!=\!0.8$ \cite{feas_anisimov_U}.
Constrained random-phase-approximation (RPA) calculations \cite{feas_nakamura_08}
on top of the DF data lead to $U\!=\!2.2\!\div\!3.3$ and $J_H\!=\!0.3\!\div\!0.6$.
Values of 4 eV were used for $U$ in DMFT
investigations by Haule {\it et al.} \cite{feas_haule_08} and Craco {\it et al.}
\cite{feas_craco_08}.
A value of $U\!=\!0.3$ eV was employed by Korshunov and Eremin \cite{feas_eremin_U}
in RPA calculations for the spin response in the  normal state of Fe pnictide
compounds.

In Ref.~\cite{feas_kroll_08}, Kroll {\it et al.} found that model-Hamiltonian 
multiplet calculations with $U\!=\!1.5$ and $J_H\!=\!0.8\!\div\!0.9$ reproduce well 
the x-ray absorption Fe $L_{2,3}$-edge spectra.
Regarding the crystal-field splittings, these authors used a value of 0.25 eV
for the $t_{2g}\!-\!e_g$ energy separation, substantially lower than our
SD-MRCI result of 0.71 eV for the ($xz$,$yz$)\,$\rightarrow$\,$3z^2\!-\!r^2$
excitation energy.
In order to reproduce the experimental spectra, a larger value for the crystal-field
splitting would imply a larger value for $U$ in the model-Hamiltonian
calculations.

\begin{figure}[b]
\includegraphics*[angle=0,width=0.90\columnwidth]{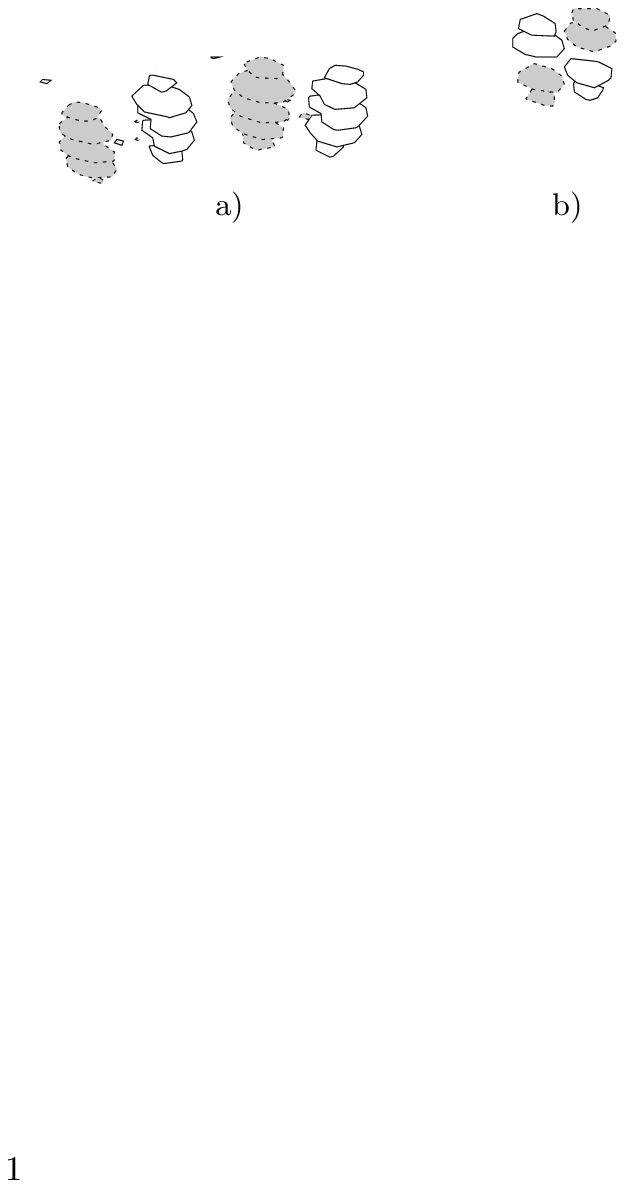}
\caption{
a) Linear combination of $4p$ hole orbitals at As sites in the $xz$ plane for the lowest
electron-removal quasiparticle state, see text.
b) Fe $3d_{xz}$  orbital for the lowest $(N\!-\!1)$ state.
}
\end{figure}

That the Coulomb interactions are substantial in these systems is best illustrated
by the nature of the lowest electron-removal states.
Our calculations show that for these states the additional holes populate the 
{\it ligand} $p$ levels, which resembles the situation in layered cuprates.

In a simple picture, the formation of oxygen $2p$ hole states in $p$-type
cuprates is due to the large Coulomb repulsion at the Cu $3d^9$ sites
\cite{ZR_88}:
in order to minimize the interaction with the Cu $3d$ holes, extra holes
in the doped system enter the O $2p$ levels.
{\it Ab intio} quantum chemical calculations show indeed that the first 
$(N\!-\!1)$ states have O $2p$ character in copper oxides, see, for
example, the analysis in Ref.~\cite{CuO_qc_08}.
The situation is quite similar in LiFeAs, where we find that for the lowest
ionized state the extra hole is accommodated into the As $4p$ orbitals.
Some details are, however, different.
Due to the $xz$ and $yz$ degeneracy, the ligand hole is distributed with equal
probability over {\it two} linear combinations of As $4p$ orbitals, i.e.,
a ``+'' combination of slightly tilted $p_x$ functions at As sites in the $xz$
plane, see Fig.~2, and a combination of tilted $p_y$ functions at As sites in the
$yz$ plane.
Not only the on-site Coulomb repulsion, but also the inter-site correlations
are effective:
in the CASSCF wave function $|\Psi\rangle$ for the lowest ionized state, the
largest weight is aquired by configurations where the two holes in the ($xz$,$yz$)
sector reside in pairs of ``orthogonal'' orbitals,
either $d_{yz}$ and $p_{x12}\!=\!p_{x1}\!+\!p_{x2}$  or
       $d_{xz}$ and $p_{y34}\!=\!p_{y3}\!+\!p_{y4}$.
Those are the first two terms in the CASSCF expansion 
$|\Psi\rangle\!=\!0.63 |p_{x12}^1 p_{y34}^2 d_{xz}^2 d_{yz}^1\rangle\!+\!
                  0.63 |p_{x12}^2 p_{y34}^1 d_{xz}^1 d_{yz}^2\rangle\!+\!
                  0.33 |p_{x12}^2 p_{y34}^2 d_{xz}^1 d_{yz}^1\rangle\!-\!
                  0.31 |p_{x12}^1 p_{y34}^1 d_{xz}^2 d_{yz}^2\rangle\!+\!...$,
where the subscripts 1,2 and 3,4 refer to As sites in the $xz$ and $yz$ planes,
respectively, and the HS-coupled electrons in the Fe $xy$, $x^2\!-\!y^2$,
and $z^2$ orbitals are omitted.
Also different from cuprates, the ligand $p$ and TM $d$ holes are HS-coupled
for the lowest $(N\!-\!1)$ state, with a total spin $S\!=\!5/2$.
Configurations where both ($xz$,$yz$)-like holes reside at the TM site
contribute as well to the $(N\!-\!1)$ wave function, see the third term in
the above expansion, as also found in cuprates for the two holes of $x^2\!-\!y^2$
symmetry \cite{CuO_qc_08,CuO_qc_09}.
In the Fe-As system, however, the two holes do not occupy the same $d$ orbital. 
The fourth term in the expression of $|\Psi\rangle$ refers to a configuration where
both holes have As $4p$ character. 

The CASSCF results for the lowest $(N\!-\!1)$ state $|\Psi\rangle$ were obtained
with a CAS that contains five Fe $d$ and two As $p$ orbitals.
The $3d$ orbitals of the Fe NN's were kept again in a $t_{2g}^6$ configuration. 
Also of interest is the nature of the next higher-lying ionized state.
That involves a hole in a composite As $4p$ orbital of $x^2\!-\!y^2$ symmetry
extending over all four ligands around the TM site, like the Zhang-Rice hole
in cuprates \cite{ZR_88}, LS-coupled to the hole in the Fe $d_{x^2-y^2}$ orbital.
The effective occupation numbers at the Fe site are
$d_{xz,yz}^3d_{xy}^1d_{3z^2-r^2}^1d_{x^2-y^2}^1$, with a HS on-site coupling,
such that the total spin for the FeAs$_4$ tetrahedron is $S\!=\!3/2$.
By CASSCF, the energy separation between the two hole states is 0.59 eV,
where the active space for the higher state $|\Psi'\rangle$ contains six orbitals.
MRCI calculations with single and double excitations from all Fe $3d$ and
As $4p$ orbitals at the FeAs$_4$ tetrahedron yield a splitting of 0.75 eV.
We also investigated the nature of the lowest electron-addition states.
Our calculations show they have Fe ($xz$,$yz$) character.

{\it Ab initio} quantum chemical calculations for determining the dispersion of
the $(N\!\mp\!1)$ quasiparticle bands and the Fermi-surface topology
of Fe pnictide systems are left for future work. 
A delicate issue is in this context the treatment of renormalization effects
due to inter-site spin interactions and spin-polaron physics.
In cuprates, we found \cite{CuO_qc_08} that such effects lead to a renormalization 
of the NN hoppings by a factor of 4.
In Fe pnictides, the experiments indicate effective mass renormalization factors 
of about 2 as compared to the DF data \cite{feas_fep_coldea_08,feas_arpes_lu_08}.

To summarize, we apply multiconfiguration CASSCF and multireference CI methods to 
the study of the multiorbital correlated electronic structure of a Fe-As compound,
the recently discovered LiFeAs.
On-site and Fe-As inter-site correlation effects are treated on equal footing in
our approach.
Our calculations predict a HS ground-state configuration for
the Fe ions, in agreement with DMFT calculations \cite{feas_haule_08,feas_craco_08}
for systems from the same family and simulations of the x-ray absorption spectra
\cite{feas_kroll_08}.
The lowest electron-removal quasiparticle states have As $4p$ character, in analogy 
to the ligand hole states in $p$-type high-$T$ cuprate superconductors.
The results indicate that the on-site Coulomb interactions are substantial.
We find that $U\!-\!U'\!\approx\!0.8$\,eV, which provides a lower bound for $U$, 
and the Hund coupling constant $J_H$ is about 0.4 eV.
Also, orbital degeneracy in the $(xz,yz)$ sector and a three-quarter filling of these
levels suggest the presence of strong fluctuations and are compatible with a
``bad-metal'' conductivity in the normal state.

We thank L. Craco for useful suggestions regarding the analysis of the 
$(N\!-\!1)$ states and for a careful reading of the manuscript.
We also acknowledge fruitful discussions with M. S. Laad and M. Gulacsi.


\begin{thebibliography}{99}
\bibitem{BS_As} The coefficients of the most diffuse $sp$ and $d$ functions for
the As basis set were increased from 0.1259 to 0.22 and from 0.407 to 0.45,
respectively.
\bibitem{footnote_RHFgs}
We do not exclude the possibility that the RHF calculation converged to a local
minimum, i.e.,  the $e_g^4d_{xy}^2$ state does not correspond to the global RHF
minimum.
\bibitem{footnote_ddHS}
SD-MRCI calculations were also performed for the lowest HS $(xz,yz)\!\rightarrow\!xy$
excited state.
To make the computations manageable, the Fe NN's were treated as closed-shell
ions, either in a $t_{2g}^6$ or $e_g^4d_{xy}^2$ configuration.
Depending on the configuration of the Fe NN's, the $(xz,yz)\!-\!xy$ CASSCF splitting
at the central site varies between 0.2 and 0.4 eV.
In each case, the SD-MRCI treatment reduces the CASSCF result by about 0.1 eV.
\bibitem{footnote_UJterms} Obviously, the Coulomb and exchange interactions are
orbital dependent in the quantum chemical calculations.
\end{thebibliography}
\end{document}